\documentclass[onecolumn,showpacs,11pt]{revtex4}

\usepackage{graphicx}
\usepackage{dcolumn}
\usepackage{bm}

\input epsf

\begin{document}

\title{\Large Reconstructing Einstein-Aether Gravity from Ordinary
and Entropy-Corrected Holographic and New Agegraphic Dark Energy
Models}

\author{\bf Ujjal Debnath\footnote{ujjaldebnath@yahoo.com ,
ujjal@iucaa.ernet.in}}

\affiliation{Department of Mathematics, Bengal Engineering and
Science University, Shibpur, Howrah-711 103, India.\\}

\date{\today}

\begin{abstract}
Here we briefly discuss the Einstein-Aether gravity theory by
modification of Einstein-Hilbert action. We find the modified
Friedmann equations and then from the equations we find the
effective density and pressure for Einstein-Aether gravity sector.
These can be treated as dark energy provided some restrictions on
the free function $F(K)$, where $K$ is proportional to $H^{2}$.
Assuming two types of the power law solutions of the scale factor,
we have reconstructed the unknown function $F(K)$ from HDE and
NADE and their entropy-corrected versions (ECHDE and ECNADE). We
also obtain the EoS parameter for Einstein-Aether gravity dark
energy. For HDE and NADE, we have shown that the type I scale
factor generates the quintessence scenario only and type II scale
factor generates phantom scenario. But for ECHDE and ECNADE, the
both types of scale factors can accommodate the transition from
quintessence to phantom stages i.e., phantom crossing is possible
for entropy corrected terms of HDE and NADE models. Finally, we
show that the models are classically unstable.
\end{abstract}


\maketitle

\section{\normalsize\bf{Introduction}}

The type Ia Supernovae and Cosmic Microwave Background (CMB)
\cite{Perlmutter,Riess} observations indicate that our universe is
presently accelerating. This acceleration is caused by some
unknown matter which has the property that positive energy density
and negative pressure is dubbed as ``dark energy'' (DE).
Observations indicate that dark energy occupies about 70\% of the
total energy of the universe, and the contribution of dark matter
is $\sim$ 26\%. Recent WMAP data analysis \cite{Briddle,Spergel}
also give us the confirmation of this acceleration. Although a
long-time debate has been done on this well-reputed and
interesting issue of modern cosmology, we still have little
knowledge about DE. The most appealing and simplest candidate for
DE is the cosmological constant $\Lambda$. Over the past decade
there have been many theoretical models for mimicking the dark
energy behaviors, such as $\Lambda$CDM, containing a mixture of
cosmological constant $\Lambda$ and cold dark matter (CDM).
However, two problems arise from this scenario, namely
``fine-tuning'' and the ``cosmic coincidence'' problems. In order
to solve these two problems, many dynamical DE theoretical models
have been proposed \cite{Cop,Arme,Sen,Feng,Kamen}. The scalar
field or quintessence \cite{Peebles,Cald} is one of the most
favored candidate of DE which produce sufficient
negative pressure to drive acceleration.\\

In recent times, considerable interest has been stimulated in
explaining the observed dark energy by the {\it holographic dark
energy} (HDE) model \cite{Enq,Zhang1,Pav}. An approach to the
problem of this dark energy arises from the holographic principle
\cite{Fis}. For an effective field theory in a box size $L$ with
UV cutoff $\Lambda_{c}$, the entropy $L^{3}\Lambda_{c}^{3}$.
Taking the whole universe into account the largest IR cut-off $L$
is chosen by saturating the inequality so that we get the
holographic dark energy density as \cite{Zhang1,Li1} in the form
$\rho_{\Lambda}=3c^{2}M_{p}^{2}L^{-2}$ where $c$ is a numerical
constant and $M_{p}\equiv1/\sqrt{8\pi G}$ is the reduced Planck
mass. The most natural choice of $L$ is the Hubble horizon
$H^{-1}$. However, Hsu \cite{Hsu} and Li \cite{Li1} pointed out
that in this case the EoS of holographic dark energy is zero and
the expansion of the universe cannot be accelerated. The next
choice of $L$ is the particle horizon. Unfortunately, in this
case, the EoS of HDE is always larger than $-1/3$ and the
expansion of the universe also cannot be accelerated. Finally, Li
\cite{Li1} found out that $L$ might be the future event horizon of
the universe. On the basis of the holographic principle, several
others have studied holographic model for dark energy
\cite{Gon,Cohen,Wei1}. Obviously, in the derivation of HDE, the
black hole (BH) entropy $S_{BH}$ plays an important role. Usually,
we know that $S_{BH}=\frac{A}{4G}$, where $A~(\sim L^{2})$ is the
area of BH horizon. Due to thermal equilibrium fluctuation,
quantum fluctuation, or mass and charge fluctuations, the BH
entropy-area relation has been modified in loop quantum gravity
(LQG) in the form \cite{Baner,Modak,Sad}
$S_{BH}=\frac{A}{4G}+\xi\ln \frac{A}{4G}+\zeta$, where $\xi$ and
$\zeta$ are dimensionless constants of order unity. Recently,
motivated by this corrected entropy-area relation in the setup of
LQG, the energy density of the entropy-corrected HDE (ECHDE) was
proposed by Wei \cite{Wei2} and also details discussed
in \cite{Sad,Jam,Kar}.\\

Based on principle of quantum gravity, Cai \cite{Cai} proposed a
new dark energy model based on the energy density. The energy
density of metric fluctuations of Minkowski-spacetime is given by
$\rho_{\Lambda}\sim M_{p}^{2}/\tilde{t}^{2}$ \cite{Maz1,Maz2}. As
the most natural choice, the time scale $\tilde{t}$ is chosen to
be the age of our universe, $T=\int_{0}^{a}\frac{da}{aH}$, where
$a$ is the scale factor of our universe and $H$ is the Hubble
parameter. Therefore, the dark energy is called {\it agegraphic
dark energy} (ADE) \cite{Cai}. The energy density of agegraphic
dark energy is given by
$\rho_{\Lambda}=3\alpha^{2}M_{p}^{2}T^{-2}$, where the numerical
factor $3\alpha^{2}$ is introduced to parameterize some
uncertainties, such as the species of quantum fields in the
universe. Since the original ADE model suffers from the difficulty
to describe the matter-dominated epoch, so Wei and Cai \cite{Wei}
have chosen the time scale $\tilde{t}$ to be the conformal time
$\eta$ instead of $T$, which is defined by $dt=ad\eta$ (where $t$
is the cosmic time), the energy density is obtained as
$\rho_{\Lambda}=3\alpha^{2}M_{p}^{2}\eta^{-2}$, which is called
{\it new agegraphic dark energy} (NADE) model \cite{Wei}. It was
found that the coincidence problem could be solved naturally in
the NADE model. The ADE models have given rise to a lot of
interest recently and have been examined and studied in details in
\cite{Wei,My}. Recently, very similar to the ECHDE model, the
energy density of the entropy-corrected NADE (ECNADE) was proposed
by Wei \cite{Wei2} and investigated in details
for acceleration of the universe \cite{Kar,Karami1,Karami2,Farooq,Male}.\\

Another approach to explore the accelerated expansion of the
universe is the modified theories of gravity. In this case cosmic
acceleration would arise not from dark energy as a substance but
rather from the dynamics of modified gravity \cite{Tsu}. Modified
gravity constitutes an interesting dynamical alternative to
$\Lambda$CDM cosmology. The simplest modified gravity is DGP
brane-world model \cite{Dvali}. The other alternative is $f(R)$
gravity \cite{An} where the Einstein-Hilbert action has been
modified. Other modified gravity includes $f(T)$ gravity, $f(G)$
gravity, Gauss-Bonnet gravity, Horava-Lifshitz gravity,
Brans-Dicke gravity, etc \cite{Ab,Lin,Yer,Noj,An1,Hora,Brans}.\\

Here we have assumed other type of modified gravity developed by
Jacobson et al \cite{Jacob,Jacob1}, known as {\it Einstein-Aether}
theory. Zlosnik et al \cite{Zlos,Zlos1} has proposed the
generalization of Einstein-Aether theory. These years a lot of
work has been done in generalized Einstein-aether theories
\cite{Gar,Linder,Barrow,Junt,Li,Gasp1,Gasp2}. In the generalized
Einstein-Aether theories by taking a special form of the
Lagrangian density of Aether field, the possibility of
Einstein-Aether theory as an alternative to dark energy model is
discussed in detail, that is, taking a special Aether field as a
dark energy candidate and it has been found the constraints from
observational data \cite{Meng1,Meng2}. Meng et al
\cite{Meng1,Meng2} have shown that only Einstein-Aether gravity
may be generated dark energy, which caused the acceleration of the
universe.\\

Recently the reconstruction procedure or correspondences between
various dark energy models became very challenging subject in
cosmological phenomena. Correspondence between different DE
models, reconstruction of DE/gravity and their cosmological
implications have been discussed by several authors \cite{Kar,
Moha,Dao,Hou,Set1,Set2,Set3,Debn1,Cho,Jaw1,Jaw2}. We shall
reconstruct of Einstein-Aether gravity model with HDE and NADE
separately. For this purpose, we first briefly discuss the
Einstein-Aether gravity theory by modification of Einstein-Hilbert
action in section II. We find the modified Friedmann equations and
then from the equations we find the effective density and pressure
for Einstein-Aether gravity sector. These can be treated as dark
energy provided some restrictions on the free function $F(K)$.
Assuming two types of the power law solutions of the scale factor,
we can reconstruct the unknown function $F(K)$ from HDE and NADE
and their entropy-corrected versions (ECHDE and ECNADE) in section
III. Finally, we give some cosmological implications of the
reconstructed models in section IV. \\

\section{\normalsize\bf{Einstein-Aether Gravity Theory and Modified Friedmann Equations}}

Einstein-Aether theory is the extension of general relativity (GR)
that incorporates a dynamical unit timelike vector field (i.e.,
Aether) coupling with the metric. The action of the
Einstein-Aether gravity theory with the normal Einstein-Hilbert
part action can be written in the form \cite{Zlos,Meng1}
\begin{equation}
S=\int d^{4}x\sqrt{-g}\left[\frac{R}{16\pi G}+{\cal L}_{EA}+{\cal
L}_{m} \right]
\end{equation}

where ${\cal L}_{EA}$ is the Lagrangian density for vector field
and ${\cal L}_{m}$ denotes the Lagrangian density for matter
field. The Lagrangian density for the vector field
\cite{Zlos,Meng1} is given by
\begin{equation}
{\cal L}_{EA}=\frac{M^{2}}{16\pi G}~F(K)+\frac{1}{16\pi
G}~\lambda(A^{a}A_{a}+1)~,
\end{equation}
\begin{equation}
K=M^{-2}{K^{ab}}_{cd}\nabla_{a}A^{c}\nabla_{b}A^{d}~,
\end{equation}
\begin{equation}
{K^{ab}}_{cd}=c_{1}g^{ab}g_{cd}+c_{2}\delta^{a}_{c}\delta^{b}_{d}+c_{3}\delta^{a}_{d}\delta^{b}_{c}
\end{equation}

where $c_{i}$ are dimensionless constants, $M$ is the coupling
constant, $\lambda$ is a Lagrangian multiplier, $A^{a}$ is a
contravariant vector and $F(K)$ ia an arbitrary function of $K$.
From (1), we get the Einstein's field equations
\begin{equation}
G_{ab}=T_{ab}^{EA}+8\pi G T_{ab}^{m}~,
\end{equation}
\begin{equation}
\nabla_{a}\left(F'{J^{a}}_{b}\right)=2\lambda A_{b}
\end{equation}
where
\begin{equation}
F'=\frac{dF}{dK}~~and~~{J^{a}}_{b}=2{K^{ad}}_{bc}\nabla_{d}A^{c}
\end{equation}
Here $T_{ab}^{m}$ is the energy momentum tensor for matter and
$T_{ab}^{EA}$ is the energy momentum tensor for the vector field
given as follows:
\begin{equation}
T_{ab}^{m}=(\rho+p)u_{a}u_{b}+pg_{ab}
\end{equation}
where $\rho$ and $p$ are respectively the energy density and
pressure of matter and $u_{a}=(1,0,0,0)$ is the fluid 4-velocity
vector and
\begin{equation}
T_{ab}^{EA}=\frac{1}{2}~\nabla_{d}\left[
\left({J_{(a}}^{d}A_{b)}-{J^{d}}_{(a}A_{b)}-J_{(ab)}A^{d}
\right)F' \right]-Y_{(ab)}F'+\frac{1}{2}~g_{ab}M^{2}F+\lambda
A_{a}A_{b}
\end{equation}
with
\begin{equation}
Y_{ab}=-c_{1}\left[
(\nabla_{d}A_{a})(\nabla^{d}A_{b})-(\nabla_{a}A_{d})(\nabla_{b}A^{d})
\right]
\end{equation}
where the subscript $(ab)$ means symmetric with respect to the
indices and $A^{a}=(1,0,0,0)$ is time-like unit vector.\\

We consider the Friedmann-Robertson-Walker (FRW) metric of the
universe as
\begin{equation}
ds^{2}=-dt^{2}+a^{2}(t)\left[\frac{dr^{2}}{1-kr^{2}}+r^{2}\left(d\theta^{2}+sin^{2}\theta
d\phi^{2}\right) \right]
\end{equation}
where $k~(=0,\pm 1)$ is the curvature scalar and $a(t)$ is the
scale factor. From equations (3) and (4), we get
\begin{equation}
K=\frac{3\beta H^{2}}{M^{2}}
\end{equation}
where $\beta$ is constant. From eq. (5), we get the modified
Friedmann equation for Einstein-Aether gravity as in the
following:
\begin{equation}
\beta\left(-F'+\frac{F}{2K}\right)H^{2}+\left(H^{2}+\frac{k}{a^{2}}\right)=\frac{8\pi
G}{3}~\rho
\end{equation}
and
\begin{equation}
\beta\frac{d}{dt}\left(HF'\right)+\left(-2\dot{H}+\frac{2k}{a^{2}}\right)=8\pi
G(\rho+p)
\end{equation}
where $H~(=\frac{\dot{a}}{a})$ is Hubble parameter. Also the
conservation equation is given by
\begin{equation}
\dot{\rho}+3\frac{\dot{a}}{a}(\rho+p)=0
\end{equation}

Let $\rho_{EA}$ and $p_{EA}$ be the effective energy density and
pressure governed by the Einstein-Aether gravity, then we may
write the equations (13) and (14) in the following form:
\begin{equation}
\left(H^{2}+\frac{k}{a^{2}}\right)=\frac{8\pi
G}{3}~\rho+\frac{1}{3}~\rho_{EA}
\end{equation}
and
\begin{equation}
\left(-2\dot{H}+\frac{2k}{a^{2}}\right)=8\pi
G(\rho+p)+(\rho_{EA}+p_{EA})
\end{equation}
and hence we obtain
\begin{equation}
\rho_{EA}=3\beta H^{2}\left(F'-\frac{F}{2K} \right)
\end{equation}
and
\begin{equation}
p_{EA}=-3\beta H^{2}\left(F'-\frac{F}{2K}
\right)-\beta(\dot{H}F'+H\dot{F}')
\end{equation}

The equation of state (EoS) parameter due to the Einstein-Aether
contribution is given by
\begin{equation}
w_{EA}=\frac{p_{EA}}{\rho_{EA}}=-1- \frac{(\dot{H}F'+H\dot{F}')}{3
H^{2}\left(F'-\frac{F}{2K} \right)}
\end{equation}

Since density is always positive, so $\rho_{EA}>0$ implies
$F'>\frac{F}{2K}$~, where we assume $\beta>0$. The effective
density and pressure governed by Einstein-Aether gravity generate
dark energy if $\rho_{EA}+3p_{EA}<0$ (i.e., strong energy
condition violates), which provides the condition
$2H^{2}\left(F'-\frac{F}{2K} \right)>-(\dot{H}F'+H\dot{F}')$.

\section{\normalsize\bf{Reconstruction of Einstein-Aether Gravity Model}}

Since Einstein-Aether theory is the modified gravity theory and
this may generate dark energy. So this gravity can be corresponds
with other well-known dark energy models. For this purpose, we
consider the dark energy models which are holographic dark energy
(HDE) and new agegraphic dark energy (NADE) and their
entropy-corrected versions (ECHDE and ECNADE). The Einstein-Aether
gravity sector contains a function $F(K)$, so equating the density
with the dark energy models, we can find $F(K)$. So $F(K)$ can be
constructed from the dark energy models. For reconstruction of
$F(K)$ in terms of $K$, we need to know the form of scale factor
$a(t)$. For this purpose, we assume two types of power-law forms
of $a(t)$ \cite{Kar} for providing the acceleration of the
universe and we reconstruct the Einstein-Aether gravity according
to the HDE, ECHDE, NADE and ECNADE models: \\

(i) {\bf Type I:} $a(t)=a_{0}t^{m}~,m>0$, where the constant
$a_{0}$ represents the present-day value of the scale factor
\cite{Noj1}. With this choice of scale factor we obtain $H$,
$\dot{H}$ and $K$ in the form: \\
\begin{equation}
H=\frac{m}{t}~,\dot{H}=-\frac{m}{t^{2}}~,~K=\frac{3\beta
m^{2}}{M^{2}t^{2}}
\end{equation}
We see that $\dot{H}<0$, so this model corresponds to only
quintessence dominated universe (not phantom) and hence the scale
factor may be called the quintessence scale factor.\\

(ii) {\bf Type II:} $a(t)=a_{0}(t_{s}-t)^{-n}~,~t<t_{s},~n>0$,
where the constant $a_{0}$ represents the present-day value of the
scale factor, $t_{s}$ is the probable future singularity finite
time \cite{Noj1,Sad1}.
\begin{equation}
H=\frac{n}{t_{s}-t}~,~\dot{H}=\frac{n}{(t_{s}-t)^{2}}~,~K=\frac{3\beta
n^{2}}{M^{2}(t_{s}-t)^{2}}
\end{equation}
We see that $\dot{H}>0$, so this model corresponds to only phantom
dominated universe (not quintessence) and hence the scale factor
may be called the phantom scale factor.\\

\subsection{\normalsize\bf{Reconstruction from Holographic Dark Energy (HDE) Model}}

We now suggest the correspondence between the holographic dark
energy scenario and Einstein-Aether dark energy model. The
holographic dark energy (HDE) density can be written as
\cite{Wu,Hound,Pav1}
\begin{equation}
\rho_{\Lambda}=3c^{2}L^{-2}~,~~L=R_{h}
\end{equation}
where $c$ is a constant and $R_{h}$ represents the future event
horizon. From observation, the best fit value of $c$ is
$0.818^{+0.113}_{-0.097}$~. For type I scale factor, $R_{h}$ is
defined as
\begin{equation}
R_{h}=a\int_{t}^{\infty}\frac{dt}{a}=\frac{t}{m-1}~,~m>1
\end{equation}

For type II scale factor, $R_{h}$ is defined as
\begin{equation}
R_{h}=a\int_{t}^{t_{s}}\frac{dt}{a}=\frac{t_{s}-t}{n+1}
\end{equation}

For type I scale factor, equating the energy densities (i.e.,
$\rho_{EA}=\rho_{\Lambda}$), we get reconstructed equation
\begin{equation}
\frac{dF}{dK}-\frac{F}{2K}=\frac{c^{2}(m-1)^{2}}{\beta m^{2}}
\end{equation}
which immediately gives the solution
\begin{equation}
F(K)=\frac{2c^{2}(m-1)^{2} }{\beta m^{2}}~K+A_{1}\sqrt{K}
\end{equation}

For type II scale factor, we also get similar reconstructed
equation as well as (26) and we find the similar solution
\begin{equation}
F(K)=\frac{2c^{2}(n+1)^{2} }{\beta n^{2}}~K+A_{2}\sqrt{K}
\end{equation}

Here $A_{1}$ and $A_{2}$ are integration constants. For these
solutions, the EoS for Einstein-Aether gravity can be obtained as
$w_{EA}=-1+\frac{2}{3m}$ for type I model and
$w_{EA}=-1-\frac{2}{3n}$ for type II model. We see that
$-1<w_{EA}<-\frac{1}{3}$ (i.e., quintessence) if $m>1$ for type I
model and  for $n>0$, $w_{EA}<-1$ (phantom) for type II model. The
graphs of $F(K)$ w.r.t $K$ has been drawn in figure 1 for both
type I and II models. From figure, we see that the reconstruction
function $F(K)$ increases as $K$ increases for HDE model.

\begin{figure}

\includegraphics[height=2.0in]{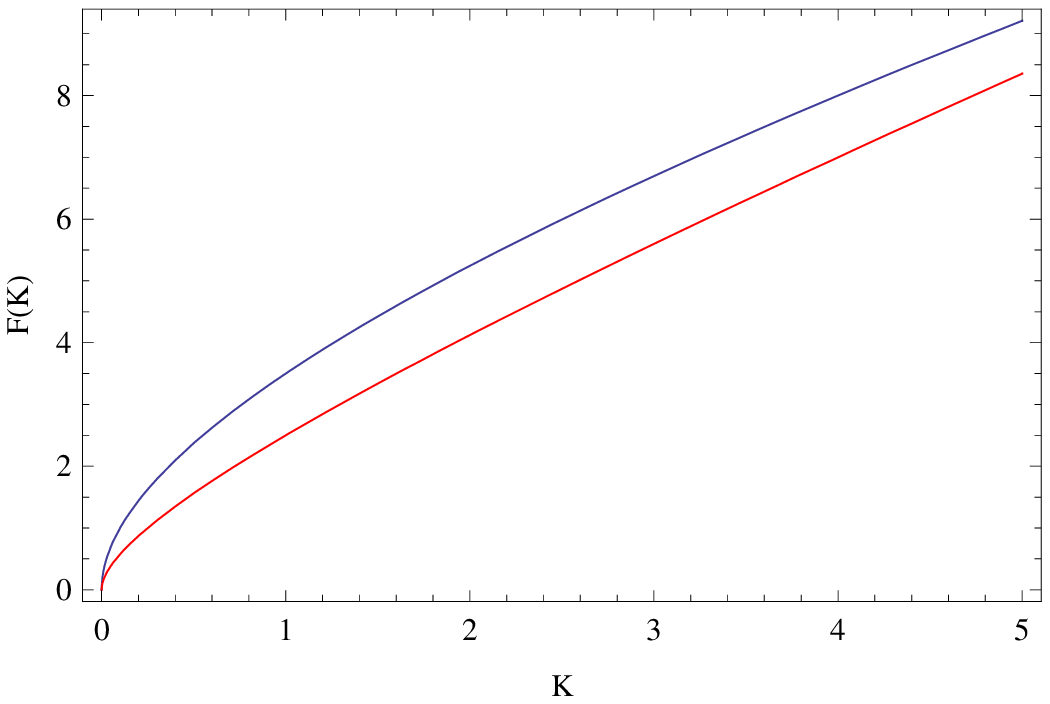}~~~~
\includegraphics[height=2.0in]{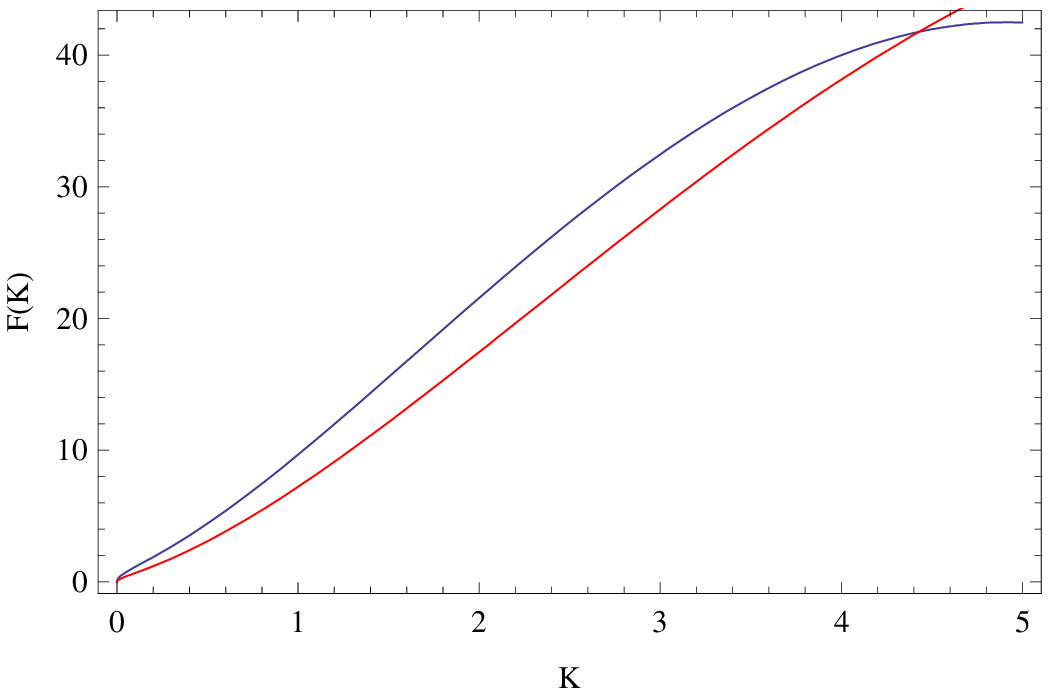}
\vspace{4mm}
~~~Fig.1~~~~~~~~~~~~~~~~~~~~~~~~~~~~~~~~~~~~~~~~~~~~~~~~~~~~~~~~~~~~~~~~~~Fig.2~~\\
\vspace{4mm}

\vspace{.2in} Figs. 1 and 2 show the variations of $F(K)$ against
$K$ for HDE and ECHDE models. Blue line represents for type I
model and red line represent for type II model. \vspace{0.2in}
\end{figure}

\subsection{\normalsize\bf{Reconstruction from Entropy-Corrected Holographic Dark Energy (ECHDE) Model}}

Using the corrected entropy-area relation, the energy density of
the ECHDE can be written as \cite{Wei2}
\begin{equation}
\rho_{\Lambda}=\frac{3c^{2}}{R_{h}^{2}}+\frac{\xi}{R_{h}^{4}}~\ln(R_{h}^{2})+\frac{\zeta}{R_{h}^{4}}
\end{equation}

where $\xi$ and $\zeta$ are constants. For type I scale factor,
equating the energy densities (i.e., $\rho_{EA}=\rho_{\Lambda}$),
we get reconstructed equation
\begin{equation}
\frac{dF}{dK}-\frac{F}{2K}=\frac{c^{2}(m-1)^{2}}{\beta m^{2}}+
\frac{(m-1)^{4}M^{2} }{(3\beta m^{2})^{2} }~K\left[\zeta+\xi
\ln\left(\frac{3\beta m^{2}}{M^{2}(m-1)^{2}K } \right) \right]
\end{equation}
which immediately give the solution
\begin{equation}
F(K)=\frac{2c^{2}(m-1)^{2} }{\beta m^{2}}~K+\frac{(m-1)^{4}M^{2}
}{9(3\beta m^{2})^{2} }~K^{2}\left[3\zeta+2\xi+
2\xi\ln\left(\frac{3\beta m^{2}}{M^{2}(m-1)^{2}K } \right)
\right]+B_{1}\sqrt{K}
\end{equation}
Here $B_{1}$ is integration constant. In this case, using eq.
(20), the EoS parameter for Einstein-Aether gravity can be written
as
\begin{equation}
w_{EA}=-1+\frac{4}{3m}\left[\frac{27c^{2}t^{2}+(m-1)^{2}\left\{9\zeta-\xi+12\xi\ln\left(\frac{t}{m-1}\right)
\right\}}{54c^{2}t^{2}+(m-1)^{2}\left\{9\zeta+2\xi+12\xi\ln\left(\frac{t}{m-1}\right)
\right\}} \right]~,m>1
\end{equation}

For type II scale factor, we also get the similar solution
\begin{equation}
F(K)=\frac{2c^{2}(n+1)^{2} }{\beta n^{2}}~K+\frac{(n+1)^{4}M^{2}
}{9(3\beta n^{2})^{2} }~K^{2}\left[3\zeta+2\xi+
2\xi\ln\left(\frac{3\beta n^{2}}{M^{2}(n+1)^{2}K } \right)
\right]+B_{2}\sqrt{K}
\end{equation}
Here $B_{2}$ is integration constant. In this case, using eq.
(20), the EoS parameter for Einstein-Aether gravity can be written
as
\begin{equation}
w_{EA}=-1-\frac{4}{3n}\left[\frac{27c^{2}(t_{s}-t)^{2}+(n+1)^{2}
\left\{9\zeta-\xi+12\xi\ln\left(\frac{(t_{s}-t)}{n+1}\right)
\right\}}{54c^{2}(t_{s}-t)^{2}+(n+1)^{2}\left\{9\zeta+2\xi+12\xi\ln\left(\frac{(t_{s}-t)}{n+1}\right)
\right\}} \right]
\end{equation}

The graphs of $F(K)$ w.r.t $K$ has been drawn in figure 2 for both
type I and II models. From figure, we see that the reconstruction
function $F(K)$ increases as $K$ increases for ECHDE model. The
EoS parameter $w_{EA}$ against time $t$ has been drawn in figure 5
for both type I and II models. We see that Einstein-Aether EoS
parameter $w_{EA}$ gives the transition from $w_{EA}>-1$ to
$w_{EA}<-1$ stages for both type I and II models. So for type I
and II models, when we assume ECHDE, the Einstein-Aether DE
interpolates from quintessence era to phantom stage. So phantom
crossing is possible for these models. Thus we conclude that for
both type I and type II models, entropy corrected terms ($\xi\ne
0,~\zeta\ne 0$) generate the phantom crossing.

\subsection{\normalsize\bf{Reconstruction from New Agegraphic Dark Energy (NADE) Model}}

The energy density of the new agegraphic dark energy (NADE) model
is given by \cite{Wei}

\begin{equation}
\rho_{\Lambda}=\frac{3\alpha^{2}}{\eta^{2}}
\end{equation}
where, $\eta=\int \frac{dt}{a(t)}$ is the conformal time and the
numerical factor $3\alpha^{2}$ serves to parameterize some
uncertainties, which include the effect of curved spacetime, some
species of quantum fields in the universe, etc. From observation,
the best fit value of $\alpha$ is $2.716^{+0.111}_{-0.109}$~.\\

For type I model, the conformal time is
\begin{equation}
\eta=\int_{0}^{t}\frac{dt}{a}=\frac{t^{1-m}}{a_{0}(1-m)}~,m<1
\end{equation}

For type II model, the conformal time is
\begin{equation}
\eta=\int_{t}^{t_{s}}\frac{dt}{a}=\frac{(t_{s}-t)^{n+1}}{a_{0}(n+1)}
\end{equation}

For type I model, equating the energy densities (i.e.,
$\rho_{EA}=\rho_{\Lambda}$), we get the reconstructed equation
\begin{equation}
\frac{dF}{dK}-\frac{F}{2K}=\frac{a_{0}^{2}\alpha^{2}(1-m)^{2}
}{\beta m^{2}}\left(\frac{3\beta
m^{2}}{M^{2}}\right)^{m}\frac{1}{K^{m}}
\end{equation}
which immediately give the solution
\begin{equation}
F(K)=\frac{2a_{0}^{2}\alpha^{2}(1-m)^{2} }{\beta
m^{2}(1-2m)}\left(\frac{3\beta m^{2}}{M^{2}}
\right)^{m}K^{1-m}+C_{1}\sqrt{K}
\end{equation}
For type II model, we get the similar reconstructed equation,
which gives the similar solution
\begin{equation}
F(K)=\frac{2a_{0}^{2}\alpha^{2}(n+1)^{2} }{\beta
n^{2}(2n+1)}\left(\frac{3\beta n^{2}}{M^{2}}
\right)^{-n}K^{n+1}+C_{2}\sqrt{K}
\end{equation}

where $C_{1}$ and $C_{2}$ are integration constants. For these
solutions, the EoS for Einstein-Aether gravity can be obtained as
$w_{EA}=-\frac{5}{3}+\frac{2}{3m}$ for type I model and
$w_{EA}=-\frac{5}{3}-\frac{2}{3n}$ for type II model. We see that
for $\frac{1}{2}<m<1$, we have $-1<w_{EA}<-\frac{1}{3}$ i.e.,
quintessence model for type I model and $w_{EA}<-\frac{5}{3}<-1$
(phantom divide) for all $n>0$ for type II model. The graphs of
$F(K)$ w.r.t $K$ has been drawn in figure 3 for both type I and II
models. From figure, we see that the reconstruction function
$F(K)$ increases as $K$ increases for NADE model.

\begin{figure}

\includegraphics[height=2.0in]{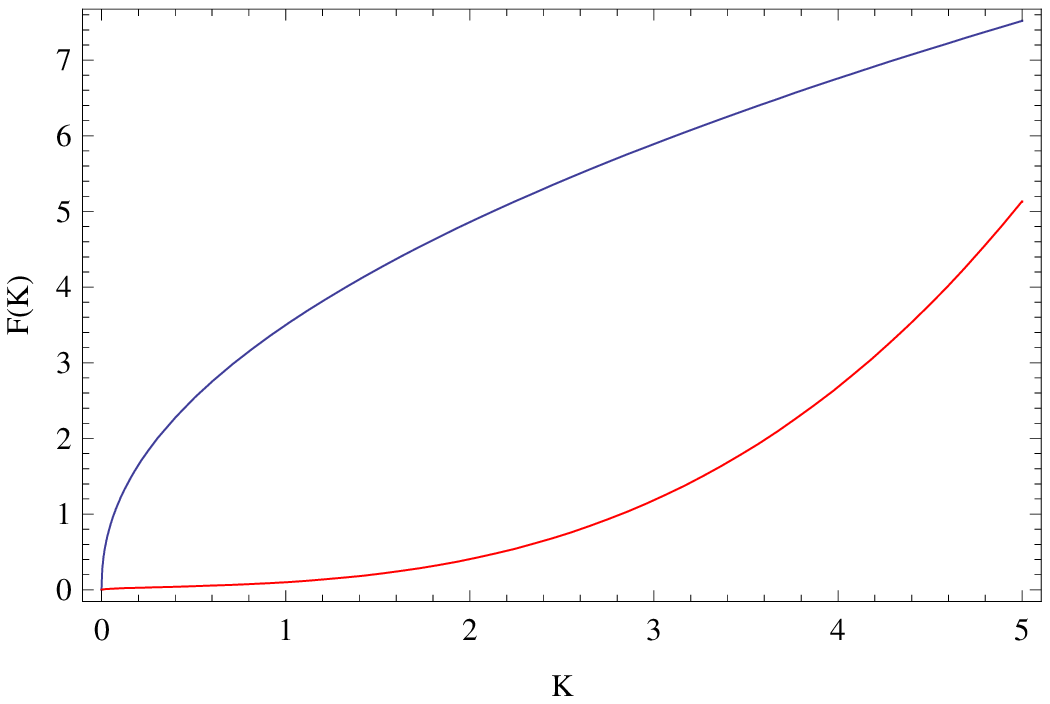}~~~~
\includegraphics[height=2.0in]{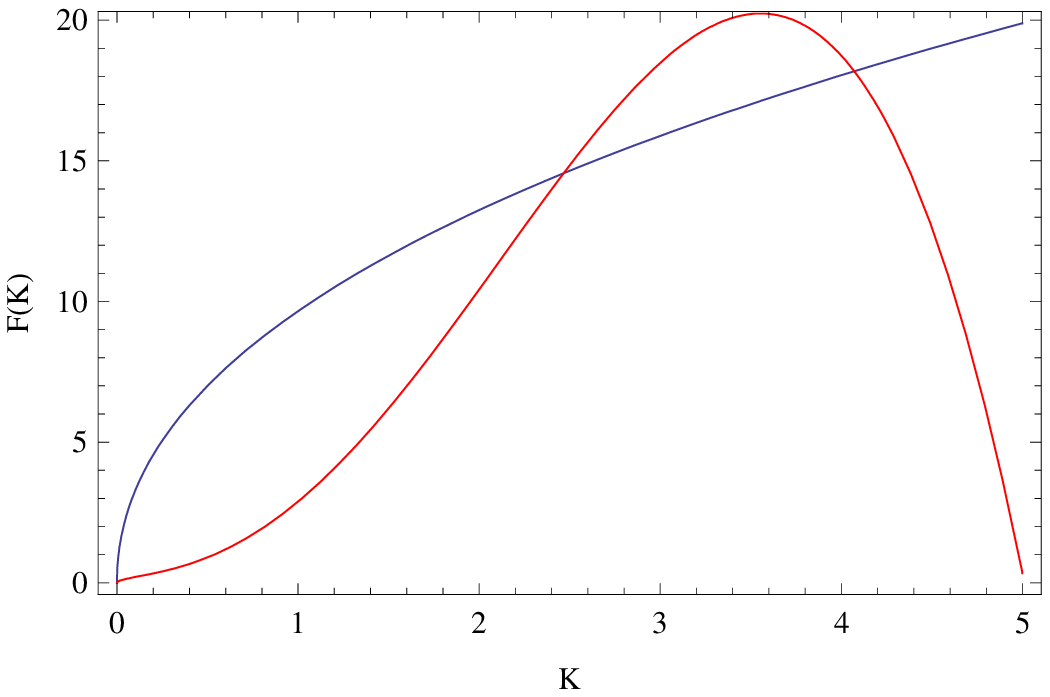}
\vspace{4mm}
~~~Fig.3~~~~~~~~~~~~~~~~~~~~~~~~~~~~~~~~~~~~~~~~~~~~~~~~~~~~~~~~~~~~~~~~~~Fig.4~~\\
\vspace{4mm}

\vspace{.2in} Figs. 3 and 4 show the variations of $F(K)$ against
$K$ for NADE and ECNADE models. Blue line represents for type I
model and red line represent for type II model. \vspace{0.2in}
\end{figure}

\subsection{\normalsize\bf{Reconstruction from Entropy-Corrected New Agegraphic Dark Energy (ECNADE) Model}}

Using the corrected entropy-area relation, the energy density of
the ECNADE can be written as \cite{Wei2}
\begin{equation}
\rho_{\Lambda}=\frac{3\alpha^{2}}{\eta^{2}}+\frac{\xi}{\eta^{4}}~\ln(\eta^{2})+\frac{\zeta}{\eta^{4}}
\end{equation}

where $\xi$ and $\zeta$ are constants. Here we have replaced
$R_{h}$ by conformal time $\eta$ in equation (29). For type I
model, equating the energy densities (i.e.,
$\rho_{EA}=\rho_{\Lambda}$), we get the reconstructed equation
\begin{eqnarray*}
\frac{dF}{dK}-\frac{F}{2K}=\frac{a_{0}^{2}\alpha^{2}(1-m)^{2}
}{\beta m^{2}}\left(\frac{3\beta
m^{2}}{M^{2}}\right)^{m}K^{-m}~~~~~~~~~~~~~~~~~~~~~~~~~~~~~~~~~~~~~~~~~~
\end{eqnarray*}
\begin{equation}
+\frac{a_{0}^{4}(1-m)^{4} }{3\beta m^{2}}\left(\frac{3\beta
m^{2}}{M^{2}}\right)^{2m-1}K^{1-2m}\left[\zeta+\xi\ln\left\{
\left(\frac{3\beta m^{2}}{M^{2}}\right)^{1-m}
\frac{K^{m-1}}{a_{0}^{2}(1-m)^{2} }\right\} \right]
\end{equation}
which immediately give the solution
\begin{eqnarray*}
F(K)=\frac{2a_{0}^{2}\alpha^{2}(1-m)^{2} }{\beta
m^{2}(1-2m)}\left(\frac{3\beta m^{2}}{M^{2}}
\right)^{m}K^{1-m}+D_{1}\sqrt{K} +\frac{2a_{0}^{4}(1-m)^{4}
}{3\beta m^{2}(3-4m)^{2} }\left(\frac{3\beta
m^{2}}{M^{2}}\right)^{2m-1}K^{2-2m}\times
\end{eqnarray*}
\begin{equation}
\times\left[(3-4m)\zeta+2(1-m)\xi +(3-4m)\xi\ln\left\{
\left(\frac{3\beta m^{2}}{M^{2}}\right)^{1-m}
\frac{K^{m-1}}{a_{0}^{2}(1-m)^{2} }\right\} \right]
\end{equation}
Here $D_{1}$ is integration constant. In this case, using eq.
(20), the EoS parameter for Einstein-Aether gravity can be written
as
\begin{equation}
w_{EA}=-1+\frac{2(1-m)}{3m}\left[\frac{a_{0}^{2}\beta
m^{2}(2\zeta-\xi)+\alpha^{2}M^{2}\left(\frac{3\beta
m^{2}}{M^{2}}\right)^{m}t^{2-2m}+4a_{0}^{2}\beta
m^{2}\xi\ln\left(\frac{t^{1-m}}{a_{0}(1-m)}\right)}
{a_{0}^{2}\beta m^{2}\zeta+\alpha^{2}M^{2}\left(\frac{3\beta
m^{2}}{M^{2}}\right)^{m}t^{2-2m}+2a_{0}^{2}\beta
m^{2}\xi\ln\left(\frac{t^{1-m}}{a_{0}(1-m)}\right)} \right]~,~m<1
\end{equation}

For type II model, we get the similar reconstructed equation,
which gives the similar solution
\begin{eqnarray*}
F(K)=\frac{2a_{0}^{2}\alpha^{2}(1+n)^{2} }{\beta
n^{2}(1+2n)}\left(\frac{3\beta n^{2}}{M^{2}}
\right)^{-n}K^{1+n}+D_{2}\sqrt{K} +\frac{2a_{0}^{4}(1+n)^{4}
}{3\beta n^{2}(3+4n)^{2} }\left(\frac{3\beta
n^{2}}{M^{2}}\right)^{-2n-1}K^{2+2n}\times
\end{eqnarray*}
\begin{equation}
\times\left[(3+4n)\zeta+2(1+n)\xi +(3+4n)\xi\ln\left\{
\left(\frac{3\beta n^{2}}{M^{2}}\right)^{1+n}
\frac{K^{-n-1}}{a_{0}^{2}(1+n)^{2} }\right\} \right]
\end{equation}
Here $D_{2}$ is integration constant. In this case, using eq.
(20), the EoS parameter for Einstein-Aether gravity can be written
as
\begin{equation}
w_{EA}=-1-\frac{2(1+n)}{3n}\left[\frac{a_{0}^{2}\beta
n^{2}(2\zeta-\xi)+\alpha^{2}M^{2}\left(\frac{3\beta
n^{2}}{M^{2}}\right)^{-n}t^{2+2n}+4a_{0}^{2}\beta
n^{2}\xi\ln\left(\frac{t^{1+n}}{a_{0}(1+n)}\right)}
{a_{0}^{2}\beta n^{2}\zeta+\alpha^{2}M^{2}\left(\frac{3\beta
n^{2}}{M^{2}}\right)^{-n}t^{2+2n}+2a_{0}^{2}\beta
n^{2}\xi\ln\left(\frac{t^{1+n}}{a_{0}(1+n)}\right)} \right]
\end{equation}
The graphs of $F(K)$ w.r.t $K$ has been drawn in figure 4 for both
type I and II models. From figure, we see that the reconstruction
function $F(K)$ increases as $K$ increases for type I model for
ECNADE. But for type II model, $F(K)$ first increases with
positive value upto a certain value of $K$ and then it sharply
decreases from positive value to negative value for increasing
$K$. The EoS parameter $w_{EA}$ against time $t$ has been drawn in
figure 6 for both type I and II models. We see that
Einstein-Aether EoS parameter $w_{EA}$ gives the transition from
$w_{EA}>-1$ to $w_{EA}<-1$ stages for both type I and II models.
So for type I and II models, when we assume ECNADE, the
Einstein-Aether DE interpolates from quintessence era to phantom
stage. So phantom crossing is possible for these models. Thus we
conclude that for both type I and type II models, entropy
corrected terms ($\xi\ne 0,~\zeta\ne 0$) generate the phantom
crossing.

\begin{figure}

\includegraphics[height=2.0in]{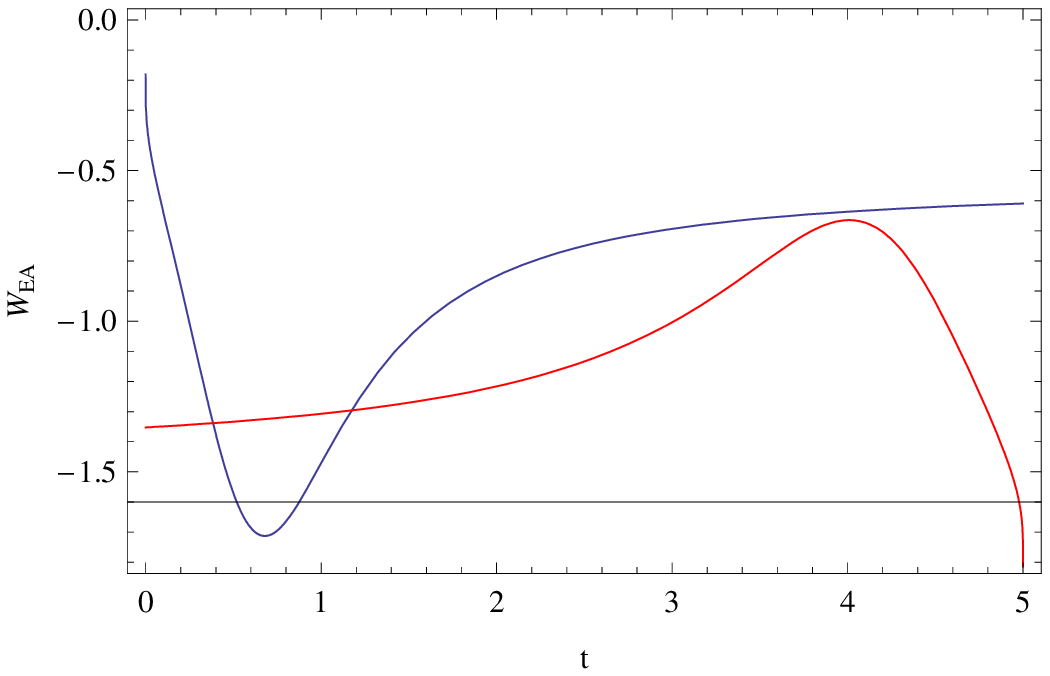}~~~~
\includegraphics[height=2.0in]{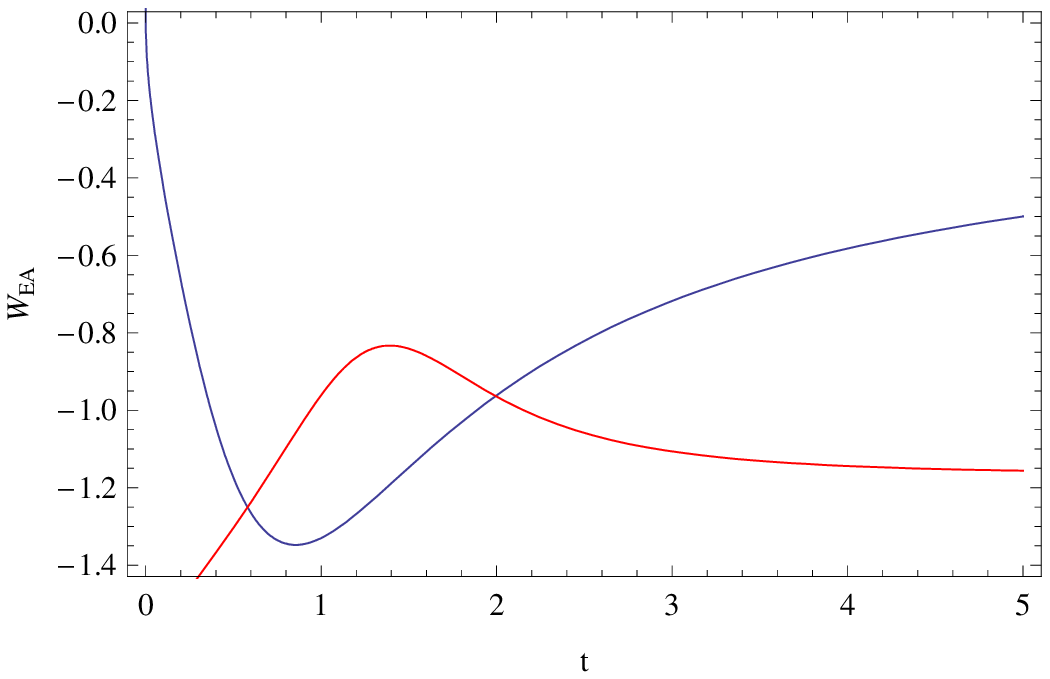}
\vspace{4mm}
~~~Fig.5~~~~~~~~~~~~~~~~~~~~~~~~~~~~~~~~~~~~~~~~~~~~~~~~~~~~~~~~~~~~~~~~~~Fig.6~~\\
\vspace{4mm}

\vspace{.2in} Figs. 5 and 6 show the variations of $w_{EA}$
against $t$ for ECHDE and ECNADE models. Blue line represents for
type I model and red line represent for type II model.
\vspace{0.2in}
\end{figure}

\section{\normalsize\bf{Discussions and Concluding Remarks}}

In this work, we have assumed the Einstein-Aether gravity theory
by modification of Einstein-Hilbert action in FRW universe. We
find the modified Friedmann equations and then from the equations
we find the effective density and pressure for Einstein-Aether
gravity sector. These can be treated as dark energy provided some
restrictions on the free function $F(K)$, where $K$ is
proportional to $H^{2}$. Assuming two types of the power law forms
of the scale factor, we have reconstructed the unknown function
$F(K)$ from HDE and NADE and their entropy-corrected versions
(ECHDE and ECNADE). From figure 1-4, we observed that the function
$F(K)$ increases with positive sign for increasing $K$ for type I
and II models when we assumed HDE, ECHDE and NADE. But for type II
model in ECNADE, $F(K)$ first increases with positive value upto a
certain value of $K$ and then it sharply decreases from positive
value to negative value for increasing $K$. We have noticed that
from the reconstructions, $F(K)\rightarrow 0$ as $K\rightarrow 0$
for all our models. We also obtain the EoS parameter for
Einstein-Aether gravity dark energy. For HDE and NADE, we have
shown that the type I scale factor generates the quintessence
scenario ($m>1$ for HDE and $\frac{1}{2}<m<1$ for NADE) only and
type II scale factor generates phantom scenario ($n>0$). So these
models cannot generate phantom crossing. But for ECHDE and ECNADE,
the EoS parameter $w_{EA}$ in terms of time $t$ have been drawn in
figures 5 and 6. For ECHDE and ECNADE, the both types of scale
factors can accommodate the transition from quintessence to
phantom stages i.e., phantom crossing is possible for these
models. Hence we conclude that phantom crossing happens for
entropy corrected terms ($\xi\ne 0,~\zeta\ne 0$) of HDE and NADE
models. Also we have observed that the forms of the constructed
function $F(K)$ and the EoS parameter $w_{EA}$ are similar for
type I and type II models of the scale
factor. \\

We may also assume the scale factor in de Sitter space time as in
the form $a(t)=a_{0}e^{Ht}$, where $H$ is constant \cite{Kar},
which can describe the early-time inflation of the universe. In
this case we shall get $K=\frac{3\beta H^{2}}{M^{2}}$, which is
constant and hence $F(K)$ must be a constant function. So the
reconstruction is not possible for de Sitter space. For this
choice of scale factor, we must have $\dot{H}=0$ and from equation
(20), we get $w_{EA}=-1$ which behaves like the cosmological
constant. From this point of view, we have noticed that Karami et
al \cite{Kar} have considered reconstruction of $f(R)$ gravity in
de Sitter space. For de Sitter space $H=$ constant implies the
Ricci scalar $R=$ constant which also implies $f(R)$ must be a
constant function. But they have obtained $f(R)$ in terms of $R$,
so the r.h.s of $f(R)$ must be constant. So for HDE, NADE, ECHDE
and ECNADE models, the $f(R)$ cannot be reconstructed in terms of
$R$. So their reconstruction analysis in de Sitter space is not correct.\\

Next we want to examine the stability of the Einstein-Aether
gravity model. For this purpose we need to verify the sign of the
square speed of sound which is defined by
$v_{s}^{2}=\frac{\partial
p_{EA}}{\rho_{EA}}=\frac{\dot{p}_{EA}}{\dot{\rho}_{EA}}$. If
$v_{s}^{2}>0$, the model is stable and $v_{s}^{2}<0$ implies the
model is classically unstable. Some authors
\cite{Jaw1,Jaw2,Kim1,My1,Eb,Sh,Setr} have shown that HDE, ADE,
NADE, Chaplygin gas, holographic Chaplygin, holographic $f(T)$,
holographic $f(G)$, new agegraphic $f(T)$, new agegraphic $f(G)$
models are classically unstable because square speed of sound is
negative throughout the evolution of the universe. In our
reconstructing Einstein-Aether gravity model from HDE, NADE,
ECHDE, ECNADE, we have to examine the signs of $v_{s}^{2}$. For
HDE model, we find $v_{s}^{2}=-1+\frac{2}{3m}$ for type I model
and $v_{s}^{2}=-1-\frac{2}{3n}$ for type II model. Since $m>1$ and
$n>0$, so we obtain $v_{s}^{2}<0$ for type I and II both models.
Also For NADE model, we find $v_{s}^{2}=-\frac{5}{3}+\frac{2}{3m}$
for type I model and $v_{s}^{2}=-\frac{5}{3}-\frac{2}{3n}$ for
type II model. Since $\frac{1}{2}<m<1$ and $n>0$, so we get
$v_{s}^{2}<0$ for type I and II both models. Hence we conclude
that the Einstein-Aether gravity for HDE and NADE models are
classically unstable. For ECHDE and ECNADE models, the expressions
of $p_{EA}$ and $\rho_{EA}$ are complicated, so we draw the graphs
of $v_{s}^{2}$ for these models in figures 7 and 8 respectively
for type I and II models. From figures, we observe that
$v_{s}^{2}<0$ for Einstein-Aether gravity in ECHDE and ECNADE
models. So our all these models are classically unstable at
present and future stages of the FRW universe.\\

\begin{figure}

\includegraphics[height=2.0in]{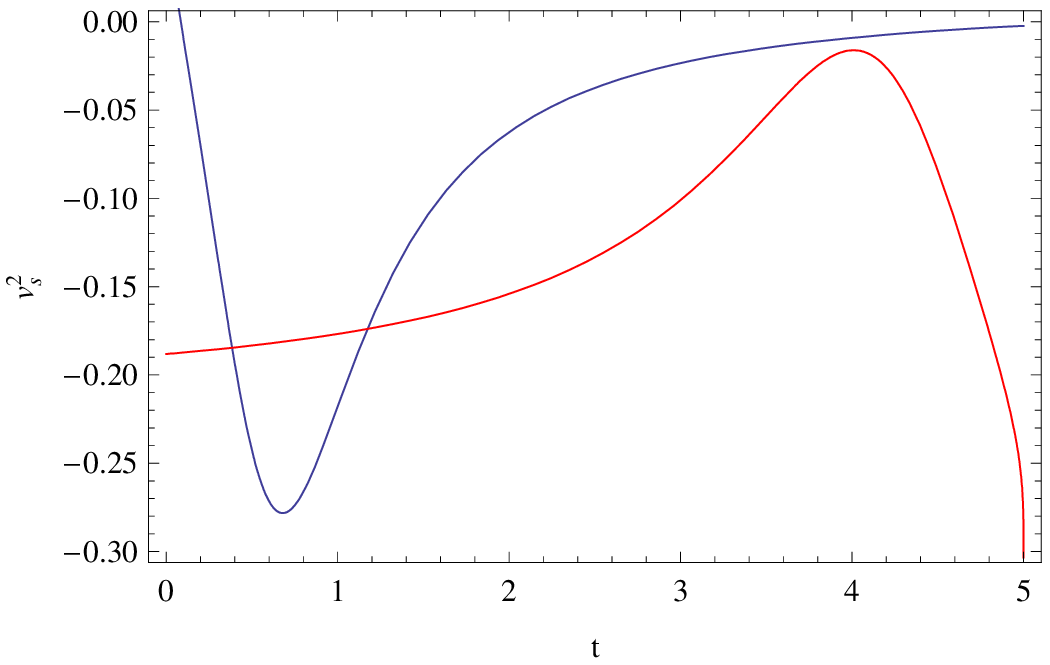}~~~~
\includegraphics[height=2.0in]{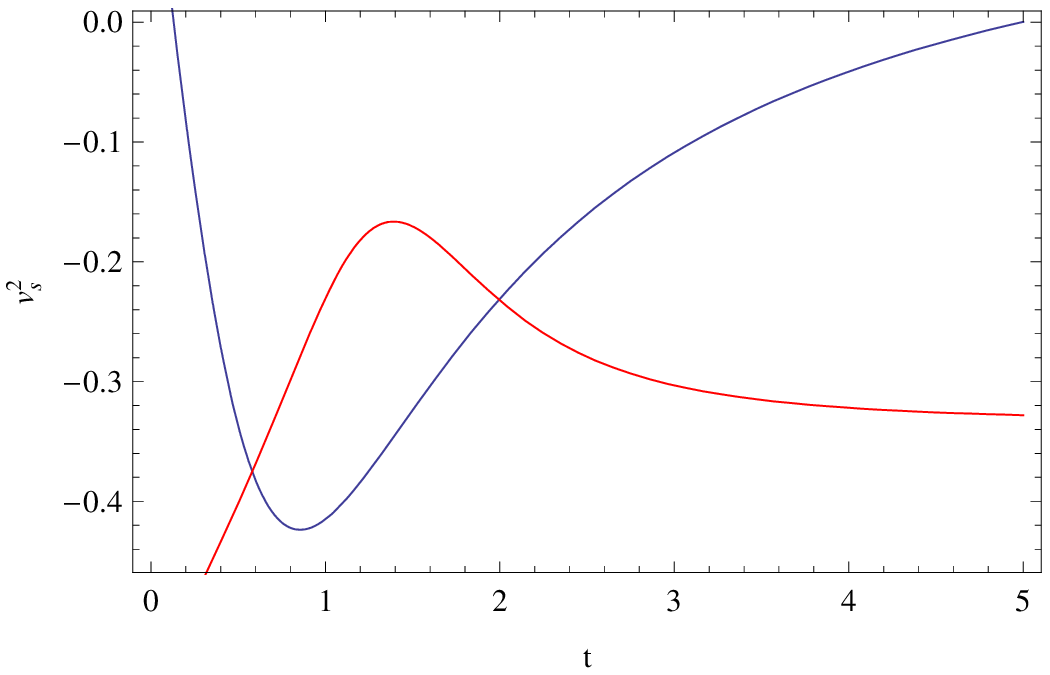}
\vspace{4mm}
~~~Fig.7~~~~~~~~~~~~~~~~~~~~~~~~~~~~~~~~~~~~~~~~~~~~~~~~~~~~~~~~~~~~~~~~~~Fig.8~~\\
\vspace{4mm}

\vspace{.2in} Figs. 7 and 8 show the variations of $v_{s}^{2}$
against $t$ for ECHDE and ECNADE models. Blue line represents for
type I model and red line represent for type II model.
\vspace{0.2in}
\end{figure}

\section*{Acknowledgements}

The author is thankful to Institute of Theoretical Physics,
Chinese Academy of Science, Beijing, China for providing TWAS
Associateship Programme under which part of the work was carried
out. Also UD is thankful to CSIR, Govt. of India for providing
research project grant (No. 03(1206)/12/EMR-II). \\

\end{document}